\documentclass[sigconf,authorversion]{acmart}

\setcopyright{acmlicensed}

\renewcommand{\paragraph}[1]{\medskip\noindent\textbf{#1:~}}
\usepackage{array}

\newcolumntype{M}[1]{>{\centering\arraybackslash}m{#1}}
\newcolumntype{N}{@{}m{0pt}@{}}
\usepackage[show]{chato-notes}
\usepackage{multirow}
\usepackage{balance}
\usepackage{siunitx}

\usepackage{tabularx}
\usepackage{makecell}
\usepackage{arydshln}
\usepackage{amsmath}
\usepackage[normalem]{ulem}
\copyrightyear{2020} 
\acmYear{2020} 
\setcopyright{acmcopyright}\acmConference[CIKM '20]{Proceedings of the 29th ACM International Conference on Information and Knowledge Management}{October 19--23, 2020}{Virtual Event, Ireland}
\acmBooktitle{Proceedings of the 29th ACM International Conference on Information and Knowledge Management (CIKM '20), October 19--23, 2020, Virtual Event, Ireland}
\acmPrice{15.00}
\acmDOI{10.1145/3340531.3411978}
\acmISBN{978-1-4503-6859-9/20/10}

\usepackage{caption}
\usepackage[labelformat=simple]{subcaption}

\newcommand{\wang}[1]{\textcolor{black}{#1}}
\newcommand{\xwang}[1]{\textcolor{black}{#1}}
\newcommand{\xiwang}[1]{\textcolor{black}{#1}}

\newcommand{\iadh}[1]{\textcolor{black}{#1}}

\newcommand{\craig}[1]{\textcolor{black}{#1}}
\newcommand{\yaya}[1]{\textcolor{black}{#1}}

\newcommand{\xw}[1]{\textcolor{black}{#1}}
\newcommand{\ya}[1]{\textcolor{black}{#1}}

\usepackage{marginnote}
\newcommand{\pageenlarge}[1]{\enlargethispage{#1\baselineskip}}
\begin{document}

%
\title[Negative Confidence-Aware Weakly Supervised Binary Classification]{Negative Confidence-Aware Weakly Supervised Binary Classification for Effective Review Helpfulness Classification}
\author{Xi Wang}
\affiliation{%
  \institution{University of Glasgow, UK}
}\email{x.wang.6@research.gla.ac.uk}

\author{Iadh Ounis}
\affiliation{%
  \institution{University of Glasgow, UK}
}\email{iadh.ounis@glasgow.ac.uk}

\author{Craig Macdonald}
\affiliation{%
  \institution{University of Glasgow, UK}
} \email{craig.macdonald@glasgow.ac.uk}

\begin{abstract}
\looseness -1 The incompleteness of positive labels and the \ya{presence of} many unlabelled instances are common problems in binary classification applications \ya{such as} in review helpfulness classification. 
Various studies from the classification literature consider all unlabelled instances as negative examples. However, a classification model that learns to classify binary instances with incomplete positive labels while assuming all unlabelled data to be negative examples will often generate a biased classifier. In this work, we propose a novel Negative Confidence-aware Weakly Supervised approach (NCWS), which customises a binary classification loss function by discriminating \ya{the} unlabelled examples with different negative \ya{confidences} during the classifier's training.
\ya{NCWS} allows to effectively, \ya{unbiasedly} identify and separate positive and negative instances after its integration into various binary classifiers from the literature, including SVM, CNN and BERT-based classifiers. 
We use the review helpfulness classification as a test case for examining the effectiveness of our \ya{NCWS} approach. We thoroughly evaluate NCWS by using three different datasets, namely one from Yelp (venue reviews), and two from Amazon (Kindle and Electronics reviews). Our results show that NCWS outperforms strong baselines from the literature including an existing SVM-based approach (i.e.\ SVM-P), the positive and unlabelled learning-based approach (i.e.\ C-PU) and the positive confidence-based approach (i.e.\ P-conf) in addressing the classifier’s bias problem. Moreover, we further examine the effectiveness of \ya{NCWS} by using its classified helpful reviews in a state-of-the-art review-based venue recommendation model  (i.e.\ DeepCoNN) and demonstrate the benefits of using NCWS in enhancing venue recommendation effectiveness in comparison to the baselines.
\pageenlarge{2}
\end{abstract}

\maketitle

\section{Introduction}\label{sec:introduction}
\looseness -1 A generic binary classifier focuses on modelling data with both positive and negative ground truth labels. However, in many binary classification applications, such as review helpfulness classification or relevant document classification, 
it is common for the ground truth to \craig{contain} only a few positive and many unlabelled instances. In particular, the unlabelled instances contain both positive and negative instances. For example, in online book reviews, only a few reviews are deemed helpful by other users (positive instances) and  many reviews are neither assessed helpful nor unhelpful by the users (unlabelled instances). This data incompleteness harms the accuracy of identifying positive and negative instances~\cite{du2015convex}. In this paper, we address the problem of classification with incomplete positive and abundant unlabelled instances. On the other hand, weakly supervised classifiers have been proposed to address classification with noisy, limited or imprecise data resources~\cite{zhou2017brief}. Indeed, binary classification with incomplete positive instances and unlabelled instances can also be addressed by a weakly supervised learning process~\cite{bao2018classification}.

\pageenlarge{2}\looseness -1 Among various weakly supervised approaches, the Positive-Unlabelled learning approach (aka PU learning) is a popular solution in addressing cases where the data has few positive instances and many unlabelled instances, by leveraging estimates of the class priors~\cite{DBLP:journals/jmlr/BlanchardLS10,DBLP:journals/jmlr/ScottB09}. However, as du Plessis~\cite{du2015convex} argued, class prior estimation-based solutions lead to a systematic estimation bias. Therefore, we propose to conduct binary weakly supervised classification on data with incomplete positive instances and unlabelled instances without the aid of an estimated class prior for the unlabelled examples. Moreover, du Plessis et al.~\cite{du2015convex} proposed to address the classifier bias problem of PU learning by applying different loss functions for the positive and unlabelled classes (an approach that we will refer to as \textbf{C-PU}). We also argue that using two customised loss functions for the positive and unlabelled data could help a classification model to fully leverage labelled and weakly labelled data, respectively. Hence, we follow du Plessis et al.~\cite{du2015convex} by using different loss functions to different classes. Meanwhile, Ishida et al.~\cite{ishida2018binary} proposed a classification approach that solely relies on the positively-labelled instances to generate a classifier according to the positivity or the confidence of the examples being positive. However, this approach filters out unlabelled examples, which can lead to \craig{a} problem of information loss and therefore negatively impacts the classification performance. 
Inspired by earlier works on PU learning~\cite{du2015convex,ishida2018binary}, we propose a negative confidence-aware weakly-supervised binary classification approach, \textbf{NCWS}, which considers \ya{both} positive and unlabelled examples with corresponding customised loss functions. In particular, we use an ordinary loss function for the positive class instead of the composite loss function $l(z) - l(-z)$ used in~\cite{du2015convex}. \xwang{Our approach is also different from~\cite{ishida2018binary}, which only uses the positive class for binary classification.} 
For the negative class, instead of using an ordinary loss function $l(z)$ for the unlabelled class, we design a customised loss function for the unlabelled class by considering the distinct probabilities of the unlabelled instances to be negative (i.e.\ negative confidence). We leverage the properties of the unlabelled instances (e.g.\ their age in the dataset since their creation time) to estimate their probabilities of being negative. As we will explain later, these properties are chosen based on their likely correlation with the negative class.
Indeed, 
our proposed NCWS approach uses \ya{additional} complementary information to the content and labels of instances (namely, the aforementioned properties of instances) to infer the likelihood of the unlabelled examples belonging to the negative class. 
 
\looseness -1 To evaluate the effectiveness of our proposed NCWS approach, we apply NCWS to \xwang{address \yaya{the} user review helpfulness classification task.}
User reviews contain various types of user opinions, including user preferences, item reputations, and item properties. However, low-quality reviews bring a certain inconvenience to users when assessing reviews and making decisions on buying a product, visiting a venue or watching a movie. It is also beneficial to business owners to be able to identify the helpfulness of reviews for their products. For example, identifying helpful reviews allows to selectively present reviews to customers, \yaya{thereby supporting them in making informed decisions}~\cite{park2018predicting}. 
To the best of our knowledge, most review helpfulness classification studies in the literature use a binary classification setup and consider reviews receiving sufficient votes\footnote{The definition of sufficiency relies on the corresponding helpfulness threshold.} as positive and the rest of reviews as negative  ~\cite{chen2016identifying,kim2006automatically}. However, reviews without helpful labels might have not yet gained views, or might have been hidden from users by the interface. \yaya{Consequently, we often observe} \xwang{a bias towards reviews that have been presented \yaya{to users}}.
\yaya{Indeed}, \yaya{the} unlabelled reviews could still be either helpful or unhelpful. Therefore, due to the prevalence of unlabelled instances in the review helpfulness classification task and the associated challenges, we argue that this task is a good \xwang{scenario} for examining the performance of our NCWS approach in addressing the weakly supervised binary classification problem. 


\looseness -1 In the review helpfulness classification task, modelling and representing reviews is key to the development of effective review helpfulness classifiers. Most existing approaches model the content of user reviews and the corresponding rating information, then make predictions on the review helpfulness labels (i.e.\ helpful or unhelpful). In this paper, we reproduce many state-of-the-art review helpfulness classification approaches as baselines and refer to these approaches as the \ya{\textit{basic}} classifiers. We select the basic classifiers with the best \ya{performances} and use them to evaluate the effectiveness of our NCWS approach. We then thoroughly examine the performance of our NCWS approach in identifying \yaya{further} helpful reviews. We extensively validate the effectiveness of our proposed NCWS approach in review helpfulness classification by investigating the extent to which accurately identifying \yaya{additional} helpful reviews can further enhance an existing review-based venue recommendation model. In particular, we use the DeepCoNN model proposed by Zheng et al.~\cite{zheng2017joint}, which is a popular \ya{state-of-the-art} review-based recommendation model.  


\pageenlarge{2}The main contributions of this paper are summarised as follows:
	\noindent\textbf{$\bullet$} We propose a weakly supervised binary classification correction approach (NCWS), which leverages positive unlabelled learning and uses a negative confidence-based loss function \ya{for} modelling \ya{the} unlabelled examples.
	
	\noindent\textbf{$\bullet$}  We \iadh{show how to integrate} NCWS within different popular binary classifiers, including \wang{SVM, CNN and BERT}-based classifiers~\cite{devlin2018bert}, \yaya{to effectively} address the review helpfulness classification task. 
	
	\noindent\textbf{$\bullet$}  We \xwang{evaluate} the effectiveness of our proposed NCWS approach by comparing it with several \yaya{existing state-of-the-art} approaches \yaya{in the literature}, including the SVM Penalty-based approach (SVM-P), a PU learning-based approach (C-PU) and a positive confidence-based approach (P-conf). 
	
	\noindent\textbf{$\bullet$}
	We \xwang{evaluate} the performance of NCWS on three real-world datasets, namely one from Yelp (Venue reviews) and two from Amazon (Kindle and Electronics reviews). 
	
	\noindent\textbf{$\bullet$}  We validate the utility of NCWS, by using its predicted helpful reviews as input to a state-of-the-art review-based recommendation model (i.e.\ DeepCoNN). 

\looseness -1 The paper is organised as follows. In Section~\ref{sec:related_work}, we describe related work. We state the tackled problem and the methodology underpinning our NCWS approach in Section~\ref{sec:methodology}. In Section~\ref{sec:baselines}, we list a number of review helpfulness \ya{classifiers}. Next, in Section~\ref{sec:experiment}, we introduce our research questions and our three used Yelp and Amazon-based datasets. We also describe the experimental setup for the basic classifiers, the baseline approaches and our NCWS approach, as well as the \ya{used} evaluation metrics for performance comparison. In Section~\ref{sec:results_analysis}, we analyse the results from the experiments to answer the research questions. In Section~\ref{sec:rq3}, we demonstrate the utility of our NCWS approach by improving the accuracy of a venue recommendation approach using the accurately classified reviews. Finally, we provide concluding remarks in Section~\ref{sec:conclusion}.

\section{Related Work}\label{sec:related_work}
\pageenlarge{2}In this section, we describe related approaches in weakly-supervised learning and review helpfulness classification.

\subsection{Weakly-supervised Approaches}
Many studies from the classification literature typically consider unlabelled instances as negative examples~\cite{DBLP:journals/tele/LeeHL18,hu2016predicting,DBLP:conf/icdm/LiuHAY08}. However, the simple grouping of unlabelled examples into a single negative class leads to inaccurate and biased binary classifiers~\cite{du2015convex,kiryo2017positive}.
Such an inaccuracy comes from the unlabelled examples.  A number of weakly-supervised learning approaches have recently been proposed to address classification with limited labelled examples. These weakly-supervised learning approaches can be used to adjust or \textit{correct} classifiers in order to improve their performances when in the presence of many unlabelled instances. Apart from the PU learning approach introduced earlier, there exist many other techniques that address the classification task with weakly supervised examples. In the following, we summarise and discuss such approaches using the categorisation of~\cite{ishida2018binary}:
(1) \textbf{Semi-supervised classification}~\cite{chapelle2009semi}, which focuses on leveraging a small amount of labelled examples to improve the performance of unsupervised learning and requires reliable examples for both positive and negative classes; (2) \textbf{One-class classification}~\cite{khan2009survey}, which focuses on distinguishing the properties of the selected class versus other classes in multi-class classification scenarios and is mainly applied to anomaly detection~\cite{ishida2018binary}; (3) \textbf{Positive-unlabelled classification}~\cite{elkan2008learning}, which is frequently adopted to address the problem of insufficient labelled examples of one class in binary classification. This approach is particularly related to our scenario consisting of binary classification with limited positive instances and unlabelled data. However, most of the positive-unlabelled classification approaches require an extra class prior estimation of the positive and unlabelled classes, which leads to a systematic estimation error~\cite{du2015convex}; (4) \textbf{Label-proportion classification}~\cite{quadrianto2009estimating}, which conducts classification according to the known class distribution. However, the class distribution is missing in our scenario; (5) \textbf{Unlabelled-unlabelled classification}~\cite{du2013clustering}, which is akin to clustering and ignores the information of the labelled examples; (6) \textbf{Complementary-label classification}~\cite{ishida2017learning} leverages extra attributes or information that denote the unrelated pattern of the corresponding class to help conduct multi-class classification; (7) \textbf{Similar-unlabelled classification}~\cite{bao2018classification} relies on the pairwise similarity of examples in one class. However, the classification will be inaccurate if the instances were wrongly labelled; (8) \textbf{Positive-confidence classification}~\cite{ishida2018binary} conducts binary classification with only positive examples and trains the classifier according to positive examples with different levels of confidence. 

Our approach in this paper is similar to the positive-unlabelled and positive-confidence classification approaches. However, unlike the positive-unlabelled classification approach, our NCWS approach does not require a class prior estimation. To illustrate the importance of this difference, we use a positive-unlabelled learning-based approach (i.e.\ C-PU~\cite{du2015convex}) as a baseline to compare with our NCWS approach. In addition, unlike the positive-confidence classification approach, we instead train the classifier by leveraging the probability of the unlabelled instances to be negative. For evaluation purposes, we use the positive-confidence classification approach (i.e.\ P-conf~\cite{ishida2018binary}) as  another baseline to our NCWS approach. 
   
Note that apart from the weakly-supervised learning, Veropoulos et al.~\cite{veropoulos1999controlling} also proposed one approach that adjusts the SVM hyperplane by putting a higher penalty on misclassifying the positive class than the negative class. However, this leads to over-sensitivity towards the positive examples when applying the bias penalty strategy \yaya{as well as} an improper boundary shape \yaya{especially with} sparse positive examples~\cite{akbani2004applying}. Moreover, we argue that over-relying on the positive examples and ignoring information behind the negative class can lead to \yaya{an} inaccurate classification. Even though the penalty-based approach has these aforementioned disadvantages when doing binary classification, we still use it as another baseline in our experiments and refer to it as `SVM-P'. 
As introduced in Section~\ref{sec:introduction}, we use the review helpfulness classification as a test case to examine the effectiveness of \yaya{NCWS}.
Therefore, we now describe the recent literature on review helpfulness classification.

\subsection{Review Helpfulness \xw{C}lassification}\label{ssec:review_helpfulness_cls} 
\pageenlarge{2} In the literature, various helpfulness classification studies have investigated different representations and properties of reviews. We classify such prior studies into five categories with different feature types, including structural features, lexical features, syntactic features, metadata features, and contextual features. 

\paragraph{Structural Features} These features capture the structure and formatting of user comments. Many studies consider structural features as strong features in detecting helpful reviews. Liu et al.~\cite{liu2007low} and Lu et al.~\cite{lu2010exploiting} leveraged the average sentence length and the number of sentences; Kim et al.~\cite{kim2006automatically} explored the effectiveness of the length of comments, the percentage of question sentences and the number of exclamation marks. In general, these studies agree that the length of reviews is one of the most effective features in review helpfulness classification. 
	
\paragraph{Lexical Features}
	Lexical features analyse the words used in the comments. Kim et al.~\cite{kim2006automatically} and Tsur and Rappoport~\cite{tsur2009revrank} used the TF-IDF score of each word and each bigram as features. Moreover, many deep learning approaches use word embedding for word representations. Word embedding-based approaches have been adopted in~\cite{chen2018cross,chen2016identifying} and have been shown to have a better expressiveness than other hand-crafted features. 
	
\paragraph{Syntactic Features}
	These features capture the linguistic properties of user comments. Kim et al.~\cite{kim2006automatically} investigated the effectiveness of different syntactic tokens including the percentage of nouns, the percentage of verbs, the percentage of adjectives and the percentage of adverbs. In addition, sentiment words were considered by Yang et al.~\cite{yang2015semantic}, who obtained significant improvements compared to using simple lexical features.  
	
	\paragraph{Metadata Features}
	These features focus on the relationship between review helpfulness and user ratings. Both Kim et al.~\cite{kim2006automatically} and Huang et al.~\cite{huang2015study} found a positive correlation between review helpfulness rating and review star ratings.

	\paragraph{Contextual Features}
	\looseness -1 These features mainly focus on the behaviour of review writers as well as the connection between the review writer and readers. Huang et al.~\cite{huang2015study} investigated the historical review helpfulness ratings of the review writers and Lu et al.~\cite{lu2010exploiting} studied the influence of the connection between the review writers and readers on \yaya{the} review helpfulness ratings. They concluded that the contextual features contribute to enhance the accuracy of the review helpfulness classification. However, these contextual features are harder to obtain than the other previous features.

In this paper, we consider some representative features from the \yaya{aforementioned} listed feature sets to develop several review helpfulness classifiers, which we use as \xwang{basic baselines}. \xwang{\yaya{Moreover,} we include 
two neural network-based classifiers as \yaya{additional} basic baselines.} 
These \xwang{basic} baselines are then corrected to alleviate the incompleteness of positive labels and the many unlabelled instances using our proposed NCWS approach in comparison to the current weakly supervised SVM-P, C-PU and P-conf approaches from the literature. In the following section, we introduce a comprehensive description of our NCWS approach.

\section{Methodology}\label{sec:methodology}

\pageenlarge{2} First, we introduce the \yaya{problem of} binary classification with limited positive and abundant unlabelled instances, as well as the used notations. Then, we illustrate our proposed NCWS approach \xwang{and how we derive the loss functions for the positive and negative classes, respectively.}

\subsection{Problem Statement} \label{ss:pbmSt}
The binary classification task consists in identifying the positive instances in a corpus of instances using binary classification approaches. This problem consists of two main objects, the set of instances $X=\{x_{1}, x_{2},..., x_{N}\}$ of size $N$ and its corresponding class label set $Y = \{y_{1}, y_{2}, ..., y_{N}\}$ with label $y_{i} \in \{+1, -1\}$. Our objective is to obtain an unbiased and accurate classifier $g(x) \rightarrow \{+1, -1\}$, which can accurately identify the positive and negative instances by modelling the limited positive and abundant unlabelled instances.

In our scenario, a $-1$ label indicates that the instance is unlabelled, rather than being negative. For example, in the review helpfulness classification test case, unlabelled reviews could be the result of a number of reasons, such as when the review is not yet old enough to have gained sufficient viewers to provide it with helpful votes, or when the user interface may have not yet shown the review to users~\cite{liu2008modeling}.  For this reason, review helpfulness classification can be seen as an example of classification with limited positive instances and many unlabelled instances. Next, we define our NCWS approach, which leverages the properties of the unlabelled instances during classification. We use the number of days $d$ since the review has been posted (i.e.\ its age in days) in addition to the content of the review to infer confidence estimates about the unlabelled reviews being actually negative. We validate the reliability of using the age of reviews as an adequate instance property in Section~\ref{ssec:datasets}.

\subsection{Negative Confidence-aware Weakly Supervised approach (NCWS)}\label{sec:age_based}
\pageenlarge{2} As introduced in Section~\ref{sec:introduction}, weak supervision provides more data to the learner. In this paper, we apply weak supervision to a classifier to address the limited positive instances and the preponderance of unlabelled instances. For example, in the review helpfulness classification task, we might reasonably assume that some unlabelled newer reviews may in fact be positive, but have not yet experienced sufficient exposure to users to gain helpful votes; conversely older unlabelled reviews are less likely to be helpful. 

\looseness -1 To address the likelihood of unlabelled instances belonging to the positive class, we propose a notion of {\em negativity}, the likelihood that an unlabelled instance belongs to the (latent) negative class. In review helpfulness classification, we assume that the likely negativity of an unlabelled review increases with its age (i.e.\ the time the review has been posted). This is motivated by the assumption that an old review has a higher probability to receive helpful votes~\cite{liu2008modeling,DBLP:journals/ecr/WangWY19,DBLP:journals/tele/LeeHL18}. Therefore, the longer time that an unlabelled review has been posted, the more likely that the review will be unhelpful\footnote{We further validate the underlying assumption in our used datasets in Section~\ref{ssec:datasets}.}. Indeed, negativity is orthogonal to the notion of positivity -- used by the PU learning approach of Ishida et al.~\cite{ishida2018binary} that only uses positive examples -- which is the confidence in the positive examples actually belonging to the positive class. In contrast, our approach models the {\em unlabelled} instances -- i.e.\ {\em NCWS} -- based on properties of these instances that indicate the confidence that they are indeed negative instances. As mentioned in Section~\ref{ss:pbmSt}, in the review helpfulness classification case, we use the age of reviews in addition to their content. In the following, we use age to describe and illustrate how we generate the negativity scores during classification.

Firstly, let the negativity score $n(x) = p(y = -1|x)$ for each unlabelled review $x$ be a function of the number of days since the review has been posted ($d(x)$):
\begin{equation}\label{equ:negativity}
    n(x) = \frac{log(d(x) + 1)}{log(max(d(X)) + 2)}
\end{equation}
\looseness -1 where $max(d(X))$ indicates the age in days of the oldest review in the \xwang{set of \yaya{review} instances $X$}.
Indeed, we argue that the longer that an unlabelled review has been posted, the more likely that the review will be unhelpful. We normalise the value of review post days $d(x)$ into the range $(0, 1)$. A higher negativity score for a review denotes a larger probability that the review is unhelpful for users. Furthermore, let $\pi_{+}$ and $\pi_{-}$ indicate the class priors $p(y= +1)$ and $p(y= -1)$, respectively.

Next, our NCWS classification approach builds a classifier, $g(x)$, by minimising the binary classification risk $R(g)$. Let the generic form of a classifier's risk $R(g)$ be as follows:
\begin{equation}\label{equa:risk_function}
    R(g) = E_{p(x,y)}\Big[\ell(y\xw{,}g(x))\Big]
\end{equation}

\noindent where $E_{p(x,y)}$ indicates the expectation over $p(x, y)$ (i.e. the probability density of instance $x$ for the corresponding label $y \in \{ +1, -1\}$), while $\ell(\cdot)$ denotes the loss function of the classifier.

Similar to the usage of \textit{example positivity} in~\cite{du2015convex}, we leverage and incorporate the instance's negativity into the risk function (i.e.\ Equation~(\ref{equa:risk_function})) as follows. First, we represent the risk function with the positive and negative prior probabilities as follows:
\begin{align}
    \begin{split}\label{equ:sub_inf_1}
        R(g) &= E_{p(x,y)}[\ell(yg(x))] = \sum_{y=\pm 1} \int \ell(yg(x))p(x|y)p(y) dx \\
        & = \int \ell(g(x))p(x|y = +1)p(y = +1) dx \\ 
        & + \int \ell(-g(x))p(x|y = -1)p(y = -1) dx \\ 
        & = \pi_{+}E_{+}\Big[\ell(g(x))\Big] + \pi_{-}E_{-}\Big[\ell(-g(x))\Big]
    \end{split}
\end{align}

The sum of the posterior probabilities of the two classes can be represented by the negative class-based posterior probability:
\begin{align}
    \label{equ:sub_inf_2}
    \pi_{+}p(x| y=+1) + \pi_{-}p(x|y = -1) & = p(x,y = +1) + p(x, y = -1) \nonumber \\
         = p(x) = \frac{p(x,y = -1)}{p(y = -1|x)} &= \frac{\pi_{-}p(x|y = -1)}{n(x)}  
\end{align}

\pageenlarge{2} Therefore, we can represent the positive part with the negative-based probabilities as follows:
\begin{align}
    \begin{split}\label{equ:sub_inf_3}
        \pi_{+}p(x|y = +1) &= \frac{\pi_{-}p(x|y=-1)}{n(x)} - \pi_{-}p(x|y = -1)\\
         & = \pi_{-}p(x|y = -1)(\frac{1-n(x)}{n(x)})
    \end{split}
\end{align}
According to Equation~(\ref{equ:sub_inf_3}), we can generate the positive summand of the risk function in Equation~(\ref{equ:sub_inf_1}) as follows:
\begin{align}
    \begin{split}\label{equ:sub_inf_4}
        \pi_{+}E_{+}\Big[l(g(x))\Big] 
        & = \int \pi_{+}p(x|y = +1)\ell(g(x)) dx \\
        & = \int \pi_{-}p(x|y = -1)(\frac{1-n(x)}{n(x)})\ell(g(x)) dx \\
        & = \pi_{-}E_{-}\Big[(\frac{1-n(x)}{n(x)})\ell(g(x))\Big]
    \end{split}
\end{align}
\craig{We} then follow the strategy proposed by du Plessis et al.~\cite{du2015convex} in modelling two classes with distinct loss or risk functions. While for the positive class we retain the original risk function (Equation~\eqref{equa:risk_function}), for the negative class, we combine Equations~(\ref{equ:sub_inf_1}) and~(\ref{equ:sub_inf_4}) as follows:  
\begin{align}
    \begin{split}
        R(g) = \pi_{-}E_{-}\Big[(\frac{1-n(x)}{n(x)})\ell(g(x)) + \ell(-g(x))\Big]
    \end{split}
\end{align}

Finally, we implement the risk function with the following objective function for the positive and negative instances respectively: 
\begin{equation}\label{equa:obj}
    R(g) = 
    \begin{cases}
        min\sum_{i = 1}^{n}\Big[\ell(g(x_{i}))\Big], & \text{if}\ y_{i} \  = \ 1 \\
        min\sum_{i = 1}^{n}\Big[(\frac{1-n(x)}{n(x))})\ell(g(x)) + \ell(-g(x))\Big], & \text{otherwise}
    \end{cases} 
\end{equation}

Thus far, we have \craig{formally} introduced our NCWS approach with the aforementioned objective function (Equation~(\ref{equa:obj})). Note that this is a general definition and can be applied to various generic binary classification approaches, and moreover, to various classification tasks for which a negativity score $n(x)$ can be defined (such as the age of review for review helpfulness). In this paper, we apply NCWS to the loss function in \yaya{an} \wang{SVM \yaya{classifier as well as two other neural network-based classifiers, \craig{specifically} classifiers} based on \xwang{CNN and BERT}}. Next, we introduce the used classifiers for review helpfulness classification.

\section{Review Helpfulness Classification}\label{sec:baselines}

In the following experiments, to demonstrate the generalisation of our NCWS approach, we use two families of classifiers. One is based on SVM along a set of hand-engineered features commonly used in the literature. \wang{We also use two neural network-based classifiers, namely \craig{CNN and BERT\xwang{-based classifiers}}.} 

\pageenlarge{2} 
\paragraph{SVM with Hand-Engineered Features} We use a support vector machine (SVM) to classify users' posted reviews into helpful and unhelpful classes. Similar to Kim et al.~\cite{kim2006automatically}, we use four groups of features in the SVM-based classifiers, namely Structural, Lexicon, Syntactic, and Metadata features. Table~\ref{tab:svm-feats} lists the \xwang{11} applied features. \xwang{\yaya{To further enhance the basic SVM-based classifiers}, we add another \yaya{metadata feature, namely Age} to the features listed in Section~\ref{ssec:review_helpfulness_cls}. \yaya{Age is a feature leveraged in our proposed NCWS approach as a key property of a review. Hence, it is added to the basic classifiers to provide additional insights into the performance of NCWS.}  \yaya{Finally,  we \xwang{combine}} all the features together into a single feature (ALL) that comprehensively considers all review's information}.
\yaya{For ease of notations, we denote by `SVM-X', an SVM classifier that uses the list of features X (e.g. `SVM-LEN' denotes the SVM classifier that is based on the LEN feature while SVM-ALL is the classifier that \xwang{that considers all features).}}

\begin{table}[]
    \caption{Categorised hand-engineered features for SVM.}\vspace{-2mm}
    \centering \scriptsize
    \begin{tabular}{|l|p{6.6cm}|}
        \hline
        \multicolumn{2}{|c|}{\textbf{Structural Features}} \\
        \hline
        LEN & The number of words included in each review.\\
        NoS & The number of sentences contained in each review.\\
        ASL & Average sentence length in each review.\\
        PoQS & Percentage of question sentences in each review.\\
        \hline
        Structural & Combines all features in this structural feature category.\\
        \hline
        \multicolumn{2}{|c|}{\textbf{Lexical Feature}} \\
        \hline
        UGR & Unigram, uses TF-IDF to generate a document feature vector for each review.\\
        \hline
        \multicolumn{2}{|c|}{\textbf{Syntactic Feature}} \\
        \hline
        Syn & The percentage of nouns, adjectives and adverbs in each review \\
        \hline
        \multicolumn{2}{|c|}{\textbf{Metadata Features}} \\
        \hline
        \xwang{Rating} & The review's corresponding rating value \\
        \xwang{Rating-Norm} & \craig{Normalised rating -} the difference between the average rating of the review's user and the corresponding rating for each review. \\
        \wang{Age} & \xwang{The number of days since \yaya{the} review was posted
        (normalised \craig{using} Eq.~\eqref{equ:negativity}).}\\
        \hline
        \hline
        \wang{ALL} & \wang{\craig{This} \xwang{combines} all hand-engineered features into one integrated feature.} 
        \\
        \hline
    \end{tabular}
    \label{tab:svm-feats}\vspace{-3mm}
\end{table}

\paragraph{\wang{Neural Network (NN) Classifiers}} \wang{We use \xwang{CNN and BERT-based classifiers} to experiment with the performance of state-of-the-art classifiers using \yaya{neural-network}-based approaches to validate the effectiveness of NCWS on different classifier types. We describe the implementation \craig{details} of these two classifiers in Section~\ref{ssec:classifier_exp}. } 

\section{Experimental Setup}\label{sec:experiment}
\looseness -1 In this section, we formulate our research questions, depict the datasets we use and provide details on the experimental setup of the baselines, our NCWS approach as well as the evaluation metrics.

\subsection{Research Questions}\label{ssec:research_questions}

In the following, we evaluate the usefulness of our NCWS approach in comparison to other weakly supervised or other \yaya{classification} \xwang{correction} approaches from the literature \yaya{(the \textit{baselines}}).
First, we validate the effectiveness of the review helpfulness classification approaches introduced in Section~\ref{sec:baselines}, which are our \textit{basic} classifiers:

\textbf{RQ1:} How effective are the review helpfulness classification approaches?
(Section~\ref{ssec:results:rq1})

\medskip Second, we evaluate our proposed NCWS approach applied to those selected basic classifiers:

\textbf{RQ2:} \yaya{Can our} proposed NCWS approach outperform \yaya{other classification correction approaches on} all used datasets (i.e.\ Yelp and Amazon) and \yaya{can it generalise to classifiers with differing performances}\xwang{? (Section~\ref{ssec:results:rq2})}

\medskip Finally, we demonstrate the usefulness of the review helpfulness models attained using NCWS to the application of venue recommendation, as per our final research question:

\textbf{RQ3:} Does \ya{an} improved NCWS-based review helpfulness \ya{classifier benefit} an existing state-of-the-art review-based venue recommendation model? (Section~\ref{sec:rq3})


\subsection{Datasets}\label{ssec:datasets}


\pageenlarge{2} \looseness -1 To address RQ1 \& RQ2, we conduct experiments on three datasets \craig{from two data sources}, namely Yelp, and the Amazon. We use the top categories from each data source with the largest amount of user reviews. Such a filtering strategy alleviates data sparseness in those categories with few user reviews, focusing instead on categories with a rich user feedback. For Yelp, we use the Yelp dataset challenge round 12\footnote{\url{https://www.yelp.com/dataset/challenge}}. After that, we collect reviews from the top three categories including `restaurants', `food', and `nightlife'. \craig{For Amazon,} we use a popular public Amazon review dataset~\cite{he2016ups}, which has been adopted in prior review helpfulness studies~\cite{diaz2018modeling,malik2018analysis}. \craig{We} select reviews from two popular categories, namely `Kindle' and `Electronics' \craig{as our two Amazon datasets.}

\looseness -1 In the Amazon-based \craig{datasets}, the user information is missing, hence \craig{for} these datasets, only the \xwang{Rating} feature is present in the Metadata feature class. As a consequence, we examine the performances of \wang{13 and 12} classifiers on the Yelp and Amazon datasets, respectively. Specifically, recall that we refer to the classification approaches we \xwang{introduced in Section~\ref{sec:baselines}}
as the basic classifiers. 
For these three datasets, following~\cite{krishnamoorthy2015linguistic}, we set the review helpfulness threshold to 1, i.e.\ we consider reviews to belong to the positive class only if they have one or more helpful votes, otherwise, the reviews are regarded as unlabelled. Table~\ref{tab:dataset_summary} provides statistics for these \yaya{datasets}.

\begin{table}[tb]
    \centering
     \caption{Summary of \ya{the used} datasets.}  \label{tab:dataset_summary}\vspace{-2mm}
    \resizebox{80mm}{!}{
    \begin{tabular}{|c|c|c|c|c|}
    \hline
        \textbf{Dataset} & \textbf{\#Reviews} & \textbf{\#Helpful} & \textbf{\#Unlabelled} & \textbf{\%Helpful}\\
    \hline
        Yelp & 1,373,587 & 621,112 & 770,780 & 45.21\% \\
        Kindle & 982,619 & 407,019 & 575,600 & 41.42\% \\
        Electronics & 1,689,188 & 633,154 & 1,056,034 & 37.48\% \\
    \hline
    \end{tabular}}\vspace{-\baselineskip}
\end{table}

\begin{figure*}[tb]
\def\HMfigurewidth{49mm}
    \centering
\begin{subfigure}{\HMfigurewidth}
\centering
    \includegraphics[width=\HMfigurewidth]{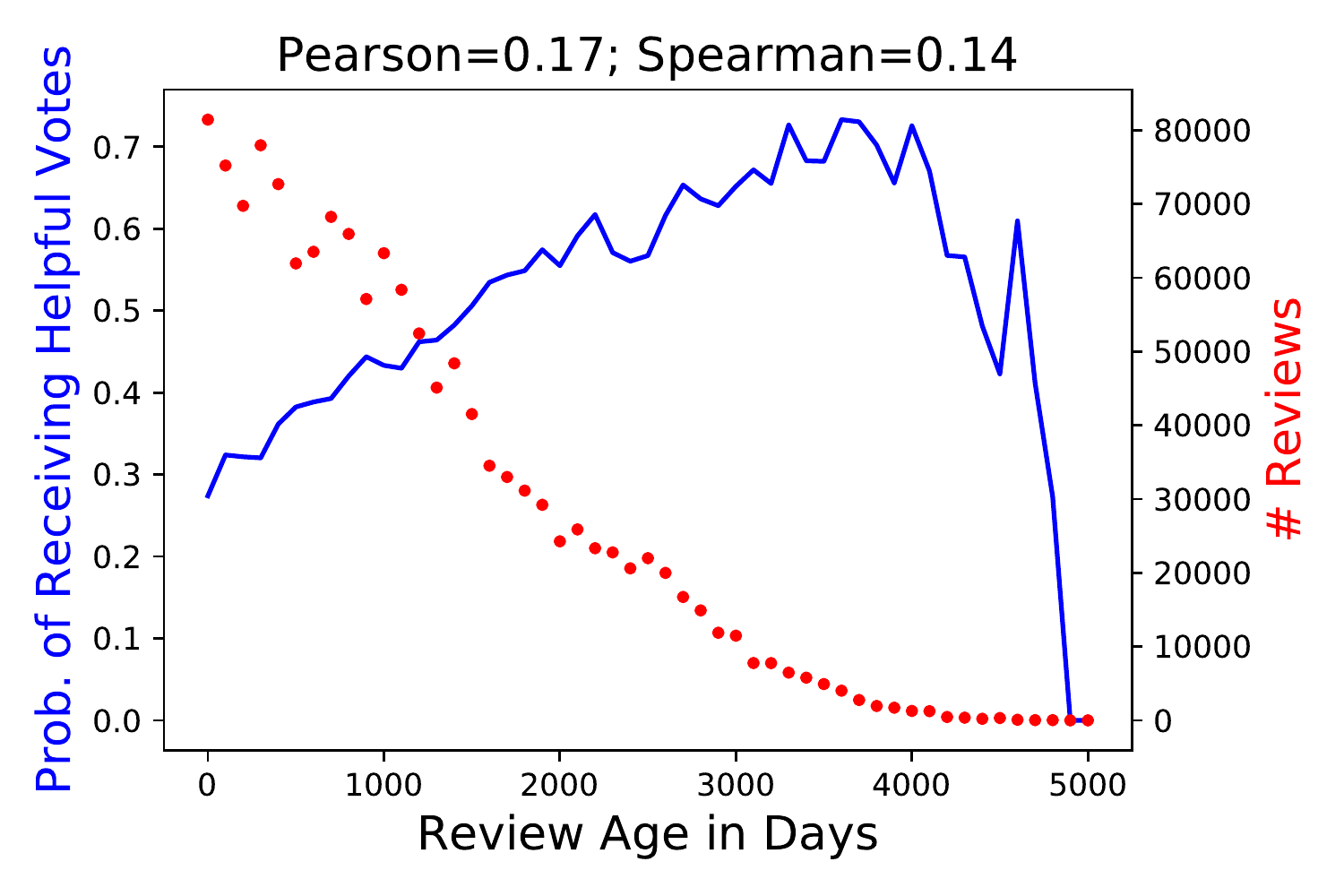}
    \caption{Yelp}
    \label{fig:Yelp_Heatmap}
\end{subfigure}
\begin{subfigure}{\HMfigurewidth}
    \centering
    \includegraphics[width=\HMfigurewidth]{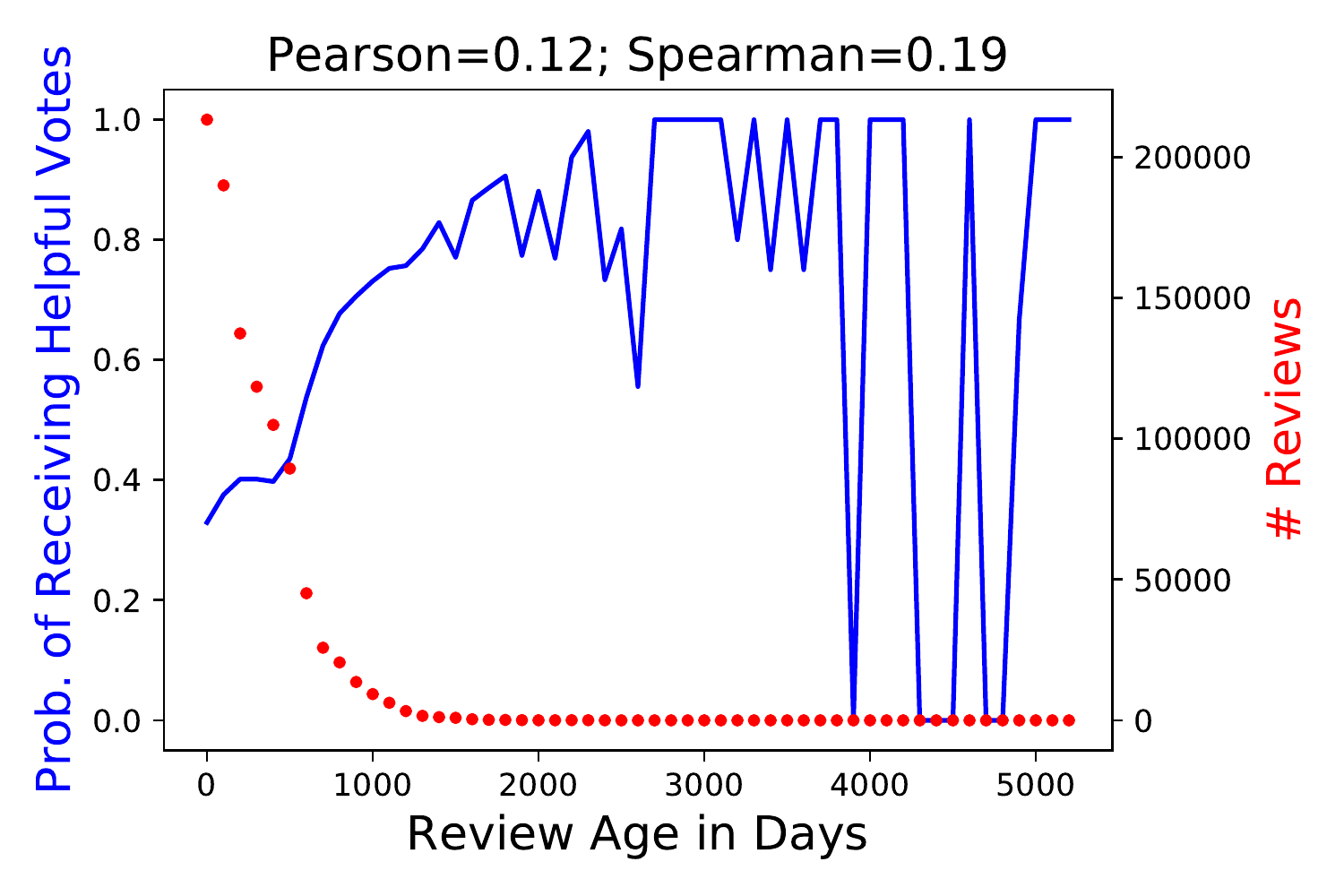}
    \caption{Kindle}
    \label{fig:kindle_Heatmap}
\end{subfigure}
\begin{subfigure}{\HMfigurewidth}
    \centering
    \includegraphics[width=\HMfigurewidth]{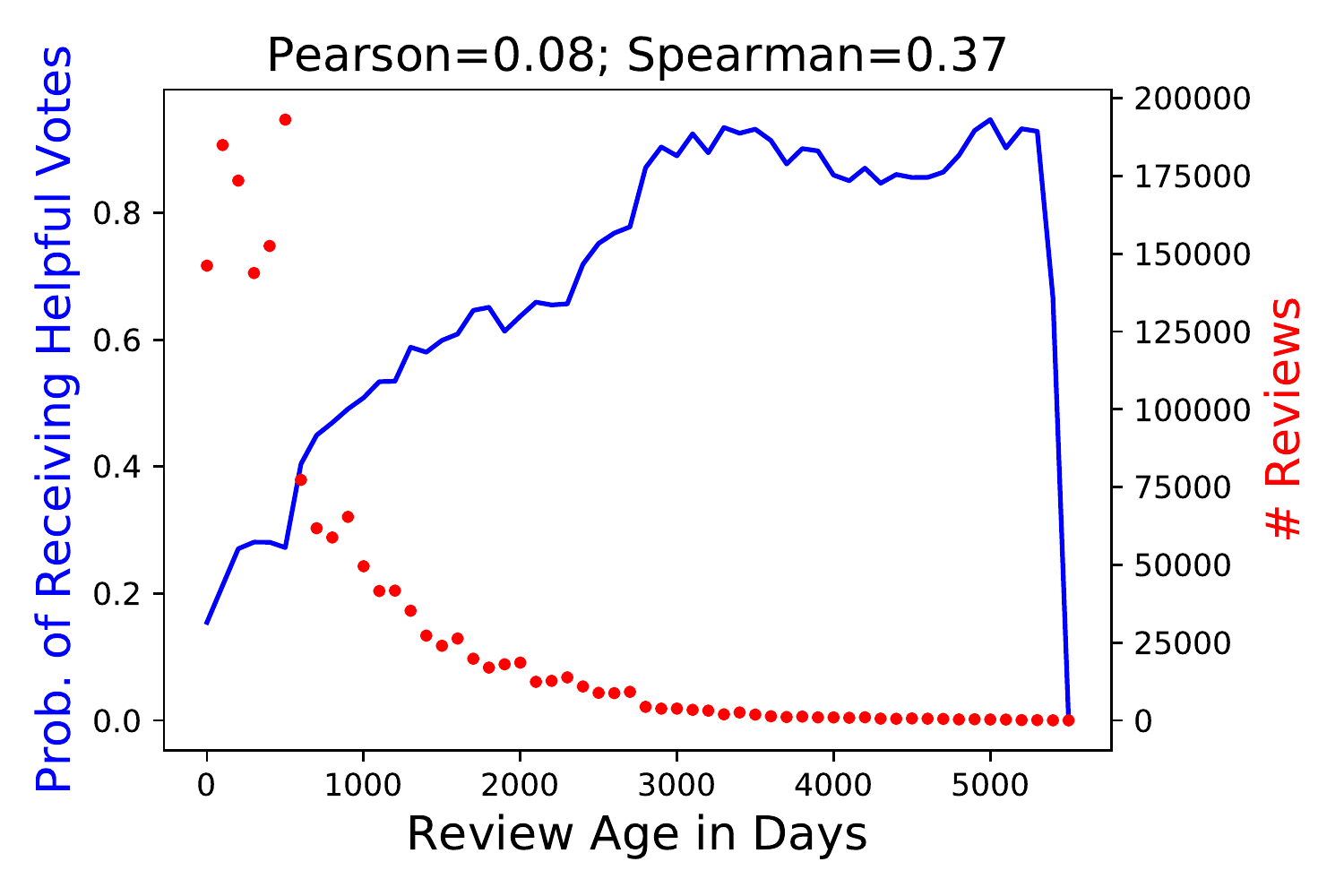}
    \caption{Electronics}
    \label{fig:electronics_Heatmap}
\end{subfigure}\vspace{-3mm}
\caption{The probability of obtaining helpful votes for reviews \ya{with} different number of days (age) since a review was posted (blue lines) and the number of posted reviews of different ages (red dots), for the Yelp, Kindle and Electronics datasets.} 
\vspace{-2mm}\end{figure*}

We conduct a 5-fold cross-validation on the three used datasets. It is apparent from the datasets summary in Table~\ref{tab:dataset_summary} that these datasets are imbalanced, having all a smaller number of helpful reviews than that of the unlabelled reviews. Therefore, we balance the class distribution of the training dataset by applying a down-sampling strategy. Down-sampling is a well-known solution to address the class distribution imbalance when training binary classifiers~\cite{sun2009classification}. In particular, as introduced in Section~\ref{sec:methodology}, we take the age of reviews into account during the calculation of the review negativity by leveraging the assumption that an older review has a higher probability to have received helpful votes. To validate our assumption on the used datasets, in Figures~\ref{fig:Yelp_Heatmap}~-~\ref{fig:electronics_Heatmap}, we plot the age of reviews (in days) against the probability that reviews of that age have been labelled as helpful. In the same figures, we also plot the review frequency by age (in red).

From Figures~\ref{fig:Yelp_Heatmap}~-~\ref{fig:electronics_Heatmap}, we observe that the Yelp, Kindle and Electronics datasets share similar helpful vote distributions across review ages. Indeed, on all three datasets, these plots appear to corroborate our assumption that there is a correlation between the number of received helpful votes by a review and the number of days the review has been posted, since the older reviews have a higher probability of receiving helpful votes than the younger reviews. In particular, we calculate the Pearson and Spearman correlations between the age of reviews (in days) and the helpful vote probability. According to the value of correlation scores, all three datasets exhibit positive Pearson and Spearman correlation scores\footnote{All correlations are statistically significant according to the scipy implementations.}. This statistically validates our assumption that older reviews have a higher probability of receiving helpful votes. Note that for the Kindle dataset, reviews that are more than 2000 days old are \xwang{rare} 
(see the corresponding frequency plot \craig{using red dots}), explaining the high variance in the corresponding helpful vote probabilities.

\pageenlarge{2} \looseness -1 Armed with these (weak) correlations, \craig{the underlying assumption of NCWS is thus:} the longer \craig{a review has been posted that remains unlabelled}, the more likely that the review will be unhelpful. \craig{\yaya{Indeed,} new reviews without votes could be helpful, but have not had sufficient opportunity to be presented to users; on the other hand, older reviews have not received helpful votes despite being presented to users.} \craig{Therefore, to evaluate the effectiveness of NCWS and its} underlying assumption, we apply NCWS to two generic types of classifiers, namely the \xwang{SVM} models with the features introduced in Section~\ref{sec:baselines} \xwang{and \yaya{the} NN-based classifiers}.
\wang{\yaya{As mentioned in Section~\ref{sec:baselines},} we also use the age of reviews as \craig{an} additional feature for the \craig{SVM classifier (denoted} SVM-Age). \yaya{Because of the \yaya{observed} correlations between the helpfulness and the age of reviews and the reliance of \xwang{NCWS} on the age property, such a \xwang{basic} baseline allows to evaluate if any improvement is the result of the use of the age feature itself}, or whether it is due to \xw{NCWS} itself.}
We \yaya{also} compare the resulting classification performances to the same classifiers but corrected using the competing methods from the literature, namely SVM-P, C-PU, and P-conf. We describe these classifiers and their corrected versions in the next section.

\subsection{Classifiers}\label{ssec:classifier_exp}

We use the \xwang{three} following so-called basic classifiers: 

\textbf{SVM:} We implement the SVM model with the LIBSVM~\cite{chang2011libsvm} library. Moreover, we use the default setting for the parameter values with a penalty parameter C = 1.0 and the RBF kernel. We instantiate different SVM classifiers based on the features sets listed in Table~\ref{tab:svm-feats}.

\textbf{CNN:} \wang{The CNN\xwang{-based} classifier consists of three vertically concatenated convolutional layers. Each layer has \craig{different} convolutional filter sizes (\craig{namely} $3\times m$, $4\times m$ and $5\times m$ respectively, where $m$ is the embedding size of a word). The output of the third convolutional layer is then \xwang{\yaya{fed into}} a linear prediction layer
to predict \yaya{the} review helpfulness label. Moreover, we use the public pre-trained word2vec vectors from Glove~\cite{pennington2014glove} with $m=100$. In particular, we adopt the cross-entropy loss to train the CNN model.}

\textbf{BERT:} \wang{We implement \yaya{a} \xwang{BERT-based classifier} with a popular natural language processing architecture (\craig{from} HuggingFace's Transformer~\cite{DBLP:journals/corr/abs-1910-03771} \craig{library}), which \craig{enables a} quick implementation of the BERT transformer to process text. In particular, we adopt the pre-trained BERT model (i.e.\ `bert-base-uncased').
Next, 
\craig{following the} setup of the CNN-based classifier, we \craig{again} use a linear prediction layer to make \yaya{the} review helpfulness \craig{predictions} and train the model by using the classic cross-entropy loss function.} 
\xwang{Both NN-based classifiers are trained using batch size 100 for 10 epochs, using the Adam optimiser with a learning rate of $10^{-4}$.}


\pageenlarge{2} \textbf{Classifier Correction Baselines:} \craig{Our experiments also apply three existing correction approaches (namely SVM-P, C-PU and P-conf ) as baselines to correct the basic classifiers:}

\paragraph{$\bullet$ SVM-P} The SVM-P approach applies a larger penalty value to the positive class than to the negative class according to the ratio of the number of positive examples versus the negative examples as suggested by Tang et al.~\cite{tang2009svms}. We use LIBSVM~\cite{chang2011libsvm} to apply the penalty values to different classes in the loss function. However, the experimental setup of SVM-P is different from other classification approaches in its training process. Indeed, the penalty values for different classes correspond to the class ratios of an unbalanced dataset. Therefore, we calculate the class ratio of the training dataset for SVM-P before down-sampling. In particular, the SVM-P approach is limited to \yaya{the} SVM-based classifiers.

\looseness -1 \paragraph{$\bullet$ C-PU} This approach was proposed by du Plessis~\cite{du2015convex}. As mentioned in Section~\ref{sec:related_work}, C-PU has a similar methodology to our NCWS approach, applying different loss functions \craig{for} the positive and unlabelled examples. However, unlike our approach, C-PU does not consider the negative confidence of unlabelled instances. For SVM, follow\-ing~\cite{du2015convex}, we use the double hinge loss, $\ell_{dh}(z) = max(-z,$ $ max(0,\frac{1}{2}-\frac{1}{2}z))$, when applying C-PU. For \craig{the} \wang{NN classifiers}, we directly integrate C-PU into \wang{the cross-entropy loss} function. 


\paragraph{$\bullet$ P-conf~\cite{ishida2018binary}} As introduced in Section~\ref{sec:related_work}, P-conf learns a classifier only from the positively-labelled instances and leverages the probability of these instances to be positive. \craig{We apply the} objective function of P-conf~\cite{ishida2018binary} to \xwang{all basic classifiers}.


\subsection{Evaluation Metrics}  
In this paper, we aim to detect positive examples with weakly supervised binary classifiers in \yaya{the} review helpfulness classification \yaya{task}. \wang{Therefore, we use the F1 metric as the key metric to evaluate the performances of the classifiers \yaya{in} accurately \yaya{classifying the} reviews. Precision and recall are \craig{also reported} to further examine the classification accuracy and the models' ability to \craig{identify} positive examples in the corpus.}


It is of note that a number of studies have \yaya{proposed approaches for the evaluation of PU learning}.  
Claesen et al.~\cite{claesen2015assessing} proposed to use a ranking-based evaluation approach and set the threshold value to divide the positive and negative examples. Jain et al.~\cite{jain2017recovering} proposed to evaluate the performance of PU learning-based classifiers with the aid of the class prior knowledge of the class distribution in the unlabelled dataset. However, the evaluation approaches of~\cite{claesen2015assessing,jain2017recovering} require \yaya{the estimation of} information such as the class threshold and the class prior, which \yaya{causes} a systematic estimation bias in the evaluation process~\cite{du2015convex}. Therefore, we \craig{resort} to using the classical evaluation metrics we introduced above and rely \yaya{only} on the ground truth of the positive examples that we have.  

\section{Results Analysis}\label{sec:results_analysis}
\looseness -1 In this section, we present and analyse the results of our experiments and answer the first two research questions in Section~\ref{ssec:research_questions}. These research questions focus on identifying the review helpfulness classifiers with the best performances among the basic classifiers and enable with higher reliability to examine the effectiveness of our proposed NCWS approach \yaya{along several classifiers of differing performances} in comparison with other classification correction approaches, namely SVM-P, C-PU and P-conf.

\subsection{RQ1: Review Helpfulness Evaluation}\label{ssec:results:rq1}\pageenlarge{2}
\looseness -1 To address RQ1 and 
\xwang{identify} 
the best performing basic classifiers, we compare the effectiveness of various \wang{basic} classifiers in distinguishing between helpful and unhelpful reviews \wang{using the F1 score}.
The results over the three used datasets \wang{are presented in Table~\ref{tab:basic_classifiers}.}

\wang{From Table~\ref{tab:basic_classifiers}}
\wang{we observe that, among the SVM-based classifiers, the SVM-LEN, SVM-Structural and SVM-ALL classifiers outperform other classification approaches and provide the best classification performances. They 
\xwang{obtain} 
the highest F1 scores across the 3 datasets ($>$0.6 on the Yelp and Electronics datasets and $>$0.45 on the Kindle dataset). Meanwhile, SVM-NoS and SVM-Age \craig{also obtain} good classification performances on the Yelp and Electronics datasets (F1 scores  $>$0.5). However, their classification performances \ya{decrease} on the \craig{Kindle} dataset ($<$0.35). In particular, by observing \yaya{the} good performances of the classifiers \craig{that deploy review length as a feature} (i.e.\ SVM-LEN, SVM-NoS, SVM-Structure and SVM-ALL), \craig{we conclude that the length of \yaya{a} review is a useful feature for predicting} review helpfulness. Furthermore, the unstable performances of the SVM-Age 
\xwang{classifier}
\craig{indicates} that the age feature \yaya{cannot by itself} fully address the review helpfulness classification problem by leveraging the (weak) correlations between the age and the helpfulness of review \craig{that was validated in Section~\ref{ssec:datasets}}. Apart from these discussed classifiers, \craig{the remainder of the} SVM-based classifiers \craig{each obtains high F1 scores on some but not \yaya{all} of the 3 datasets}
or \yaya{exhibits} bad performances across the 3 datasets. For example, \craig{the} \xwang{SVM-Rating} classifier \craig{obtains} good classification results on the Yelp and Electronics datasets but \craig{obtains} \yaya{a} very low F1 score on the Kindle dataset.}    
\wang{On the other hand, \craig{the} NN-based classifiers also provide good results with high F1 scores (>0.48). In particular, \craig{the BERT classifier obtains higher F1 scores than the CNN classifier} across \craig{all} 3 datasets. However, the best \craig{performing} SVM-based approach (i.e.\ SVM-ALL) still outperforms these two NN-based approaches on all the 3 datasets.} 

\begin{table}[]
    \centering\scriptsize
    \caption{\ya{Performances of the basic classifiers.}}\vspace{-2mm}
    \label{tab:basic_classifiers}
    \begin{tabular}{|l|c|c|c|}
        \hline
        \textbf{Basic Classifiers} & \textbf{Yelp} & \textbf{Kindle} & \textbf{Electronics} \\
        \hline
        SVM-LEN & 0.6069 & 0.4679 & 0.6047 \\
        SVM-NoS & 0.5975 & 0.3362 & 0.5759 \\
        SVM-ASL & 0.4931 & 0.0432 & 0.5315 \\
        SVM-PoQS & 0.2153 & 0.0132 & 0.5287 \\
        SVM-Structural & 0.6017 & 0.4576 & 0.6013 \\
        SVM-UGR & 0.5344 & 0.0354 & 0.0925 \\
        SVM-Syn & 0.0846 & 0.0086 & 0.0215 \\
        \xwang{SVM-Rating} & 0.5439 & 0.0753 & 0.5081 \\
        \xwang{SVM-Rating-Norm} & 0.5843 & - & - \\
        SVM-Age & 0.5428 & 0.3037 & 0.5451 \\
        SVM-ALL & \textbf{0.6340} & \textbf{0.5877} & \textbf{0.6336} \\
        \hline
        CNN & 0.5018 & 0.4830 & 0.5103 \\    
        BERT & 0.6119 & 0.5618 & 0.5712 \\
        \hline
    \end{tabular} \vspace{-3mm}
\end{table}

    

\begin{figure*}[tb]
\def\figurewidth{35mm}
\centering
\begin{subfigure}{\figurewidth}
    \includegraphics[width=\figurewidth]{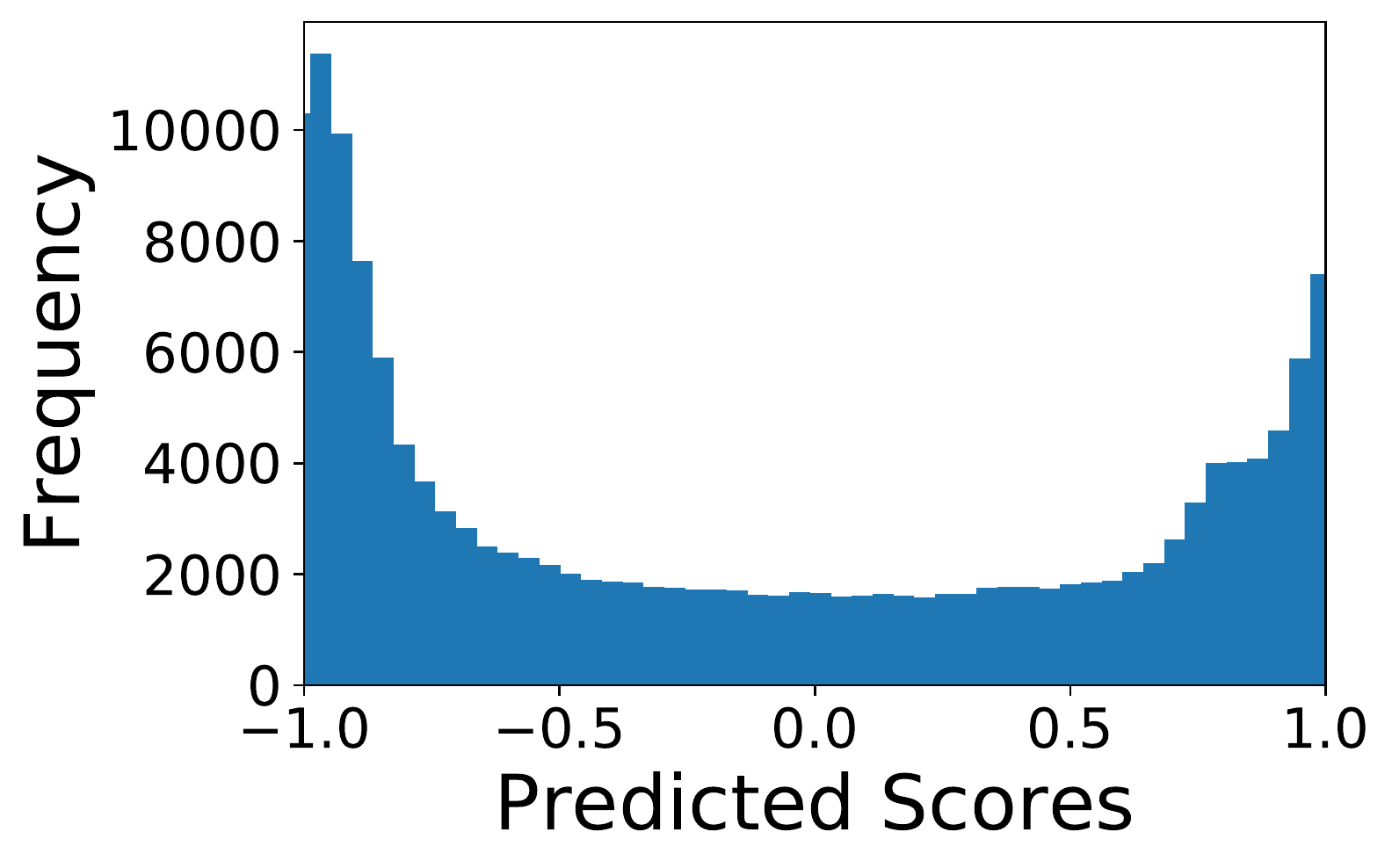}
    \caption{\craig{SVM-ALL} basic classifier}
    \label{fig:basic}
\end{subfigure}
\begin{subfigure}{\figurewidth}
    \includegraphics[width=\figurewidth]{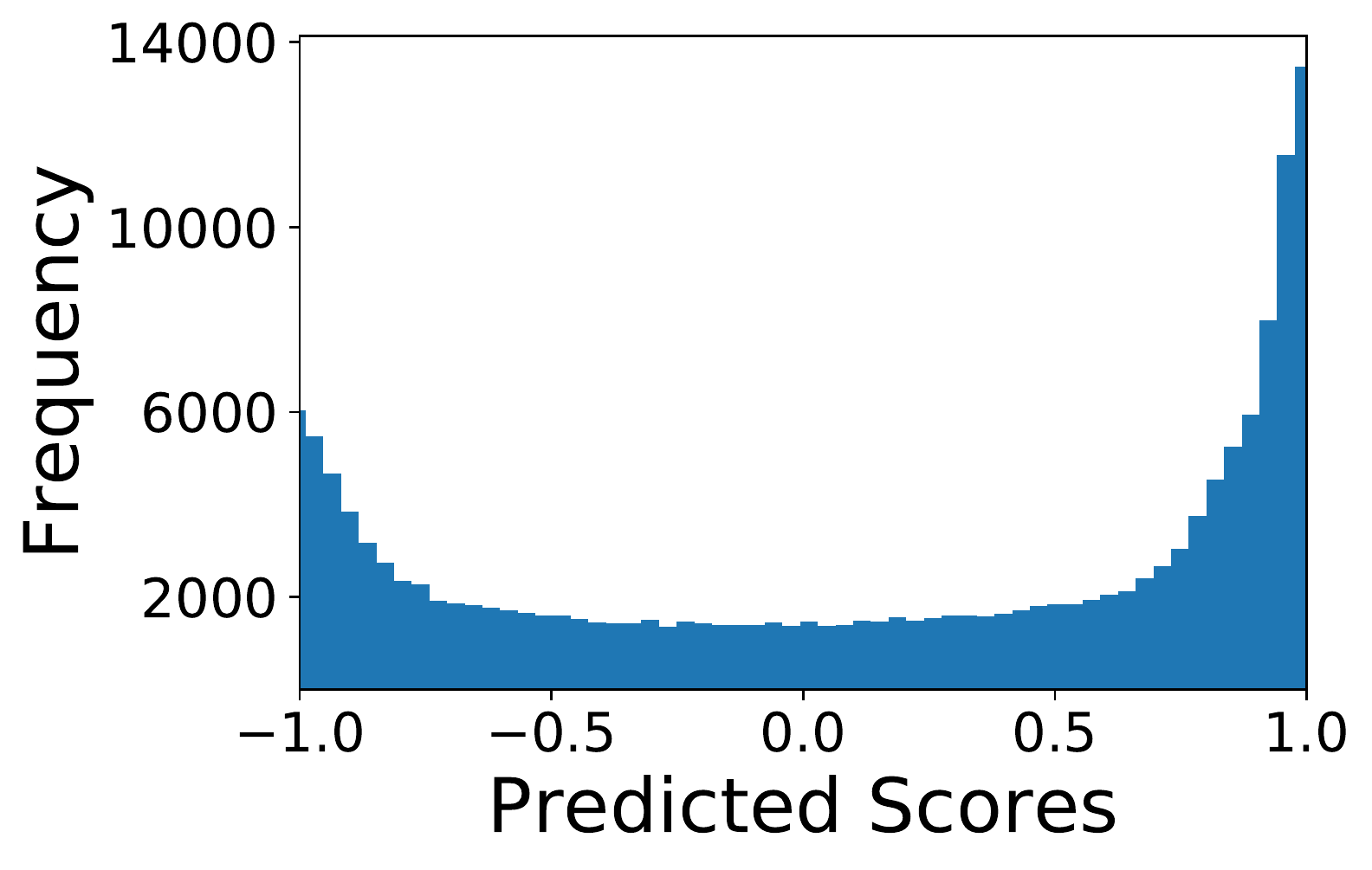}
    \caption{NCWS approach}
    \label{fig:NCWS}
\end{subfigure}
\begin{subfigure}{\figurewidth}
    \includegraphics[width=\figurewidth]{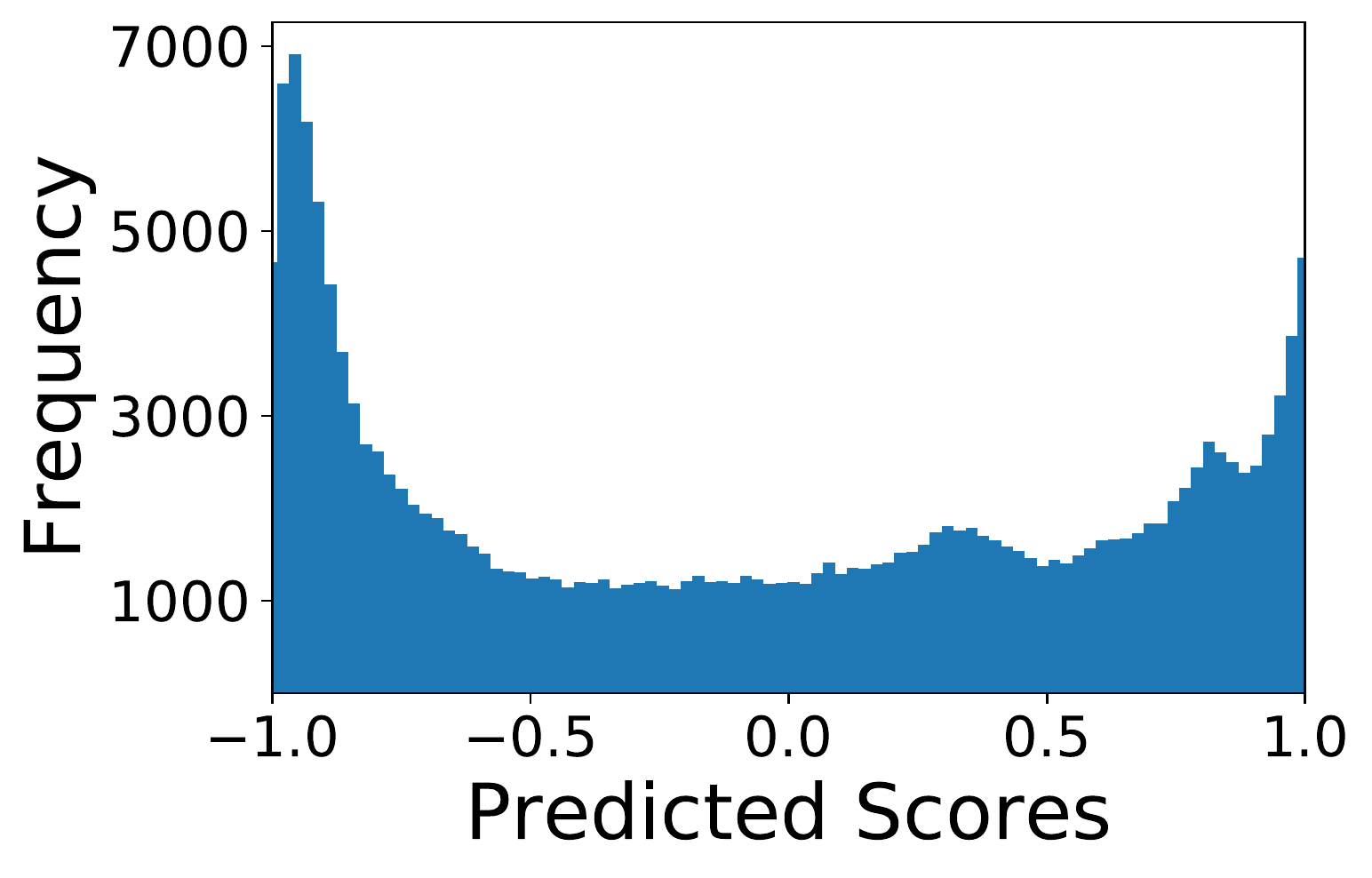}
    \caption{SVM-P approach}
    \label{fig:SVM-P}
\end{subfigure}
\begin{subfigure}{\figurewidth}
    \includegraphics[width=\figurewidth]{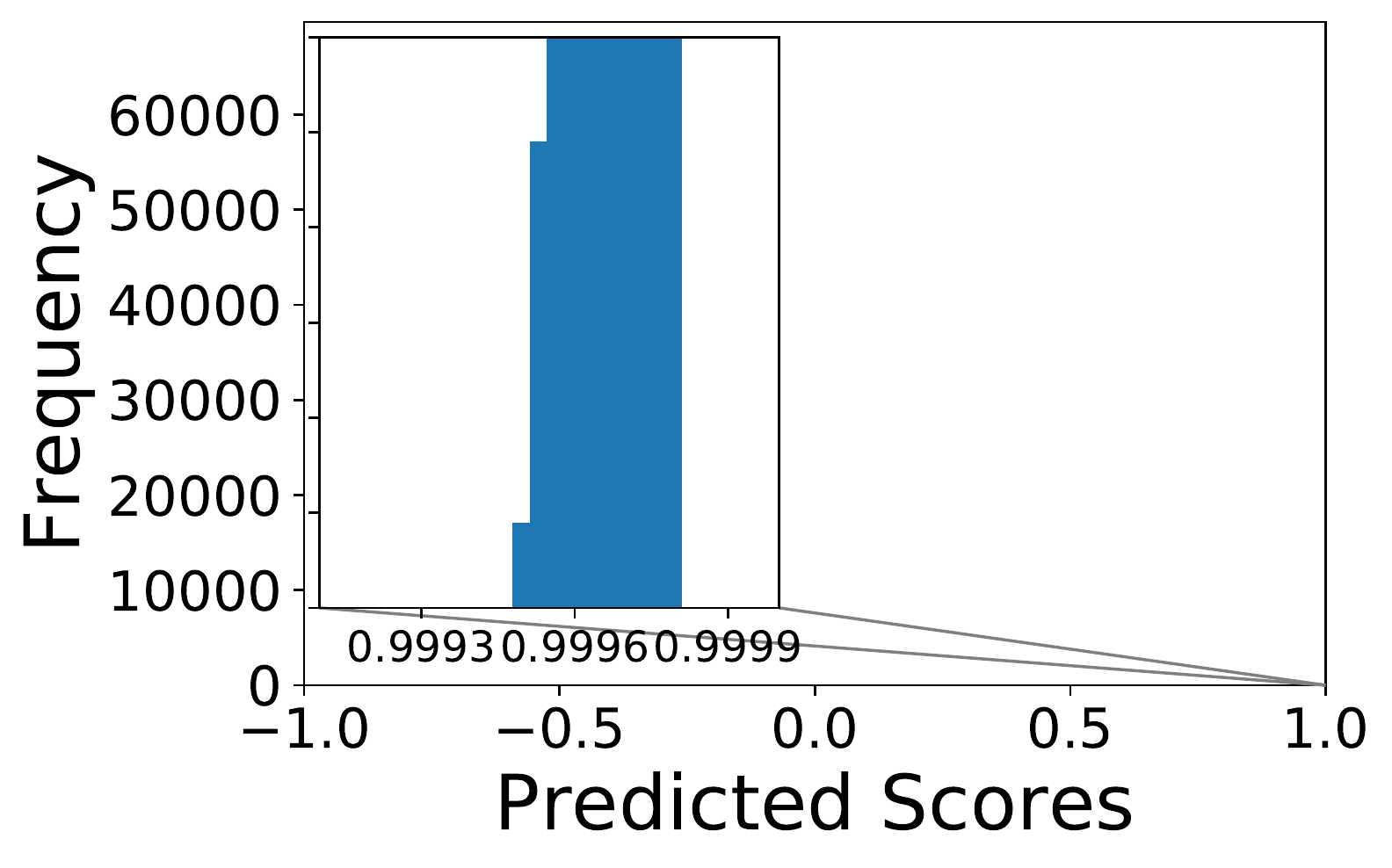}
    \caption{C-PU approach}
    \label{fig:C-PU}
\end{subfigure}
\begin{subfigure}{\figurewidth}
    \includegraphics[width=\figurewidth]{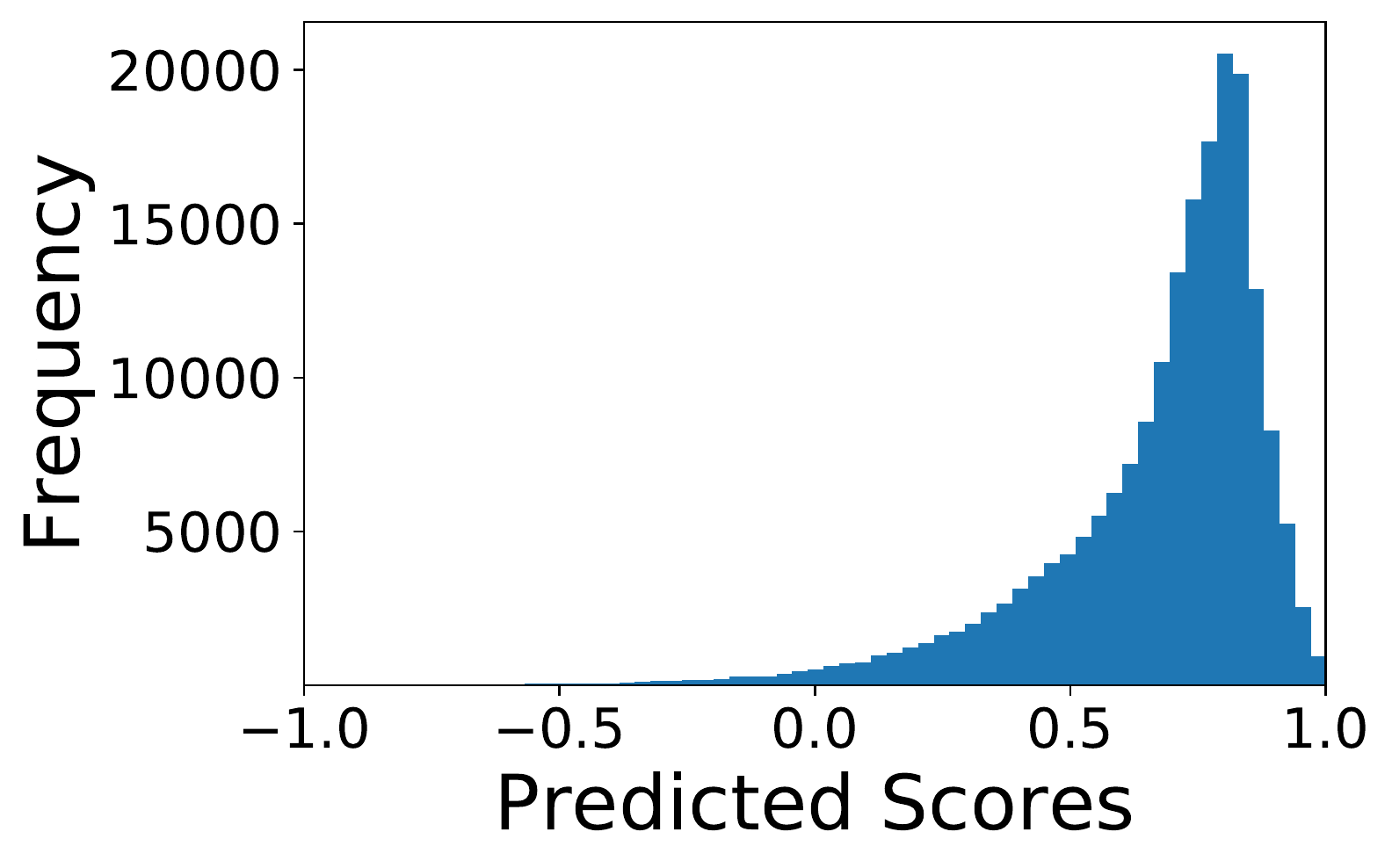}
    \caption{P-conf approach}
    \label{fig:P-conf}
\end{subfigure}\vspace{-3.5mm}
\caption{\xwang{Frequency distribution of review helpfulness score predictions for the SVM-ALL basic classifier and the corresponding classification correction approaches applied to the SVM-ALL basic classifier.}}
\vspace{-2mm}
\label{fig:baselines}
\end{figure*}
\wang{Therefore, for RQ 1, we conclude that \yaya{the basic classifiers that account for the length of a review,} including the SVM-LEN, SVM-NoS, SVM-Structural and SVM-ALL classifiers, provide \yaya{the best overall} performances among the SVM-based classifiers. This conclusion highlights the effectiveness of taking review length into account \yaya{in} the review helpfulness classification \yaya{task}. This is \craig{in line} with their good performances reported in the literature - see Section~\ref{ssec:review_helpfulness_cls}.  Moreover, the age feature-based classifier (i.e.\ SVM-Age) also shows \yaya{reasonable} classification results, which is in line with our observed correlations between the age and the helpfulness of reviews. Furthermore, the NN-based approaches (i.e.\ \yaya{the} CNN and BERT\xwang{-based \yaya{classifiers}}) \yaya{lead} \craig{also \yaya{to} competitive} \yaya{review helpfulness classifiers}. As a consequence, \yaya{we use these} \yaya{aforementioned} SVM and NN-based classifiers \yaya{as representatives} of reasonable review helpfulness classification approaches and for evaluating the effectiveness of our proposed NCWS approach.}\pageenlarge{2} 




\begin{table*}[t]
\caption{Results of \ya{the classification correction approaches} on \ya{the} Yelp, Kindle and Electronics datasets. Statistically significant differences, according to the McNemar's test ($p<0.05$), to the corresponding basic classifier are indicated by *.}\vspace{-2mm}
\label{tab:correction_effect}
    \centering\scriptsize
    \begin{tabular}{|l|c|cccc|cccc|cccc|}
\hline
        &  & \multicolumn{4}{c|}{Yelp} & \multicolumn{4}{c|}{Kindle} & \multicolumn{4}{c|}{Electronics} \\
         &  & \textbf{Precision} & \textbf{Recall} & \textbf{F1}& & \textbf{Precision} & \textbf{Recall} & \textbf{F1}& & \textbf{Precision} & \textbf{Recall} & \textbf{F1} & \\
\hline
        \multirow{3}{*}{\textbf{SVM-LEN}} & basic & 0.5781 & 0.6421 & 0.6069 & & 0.5369 & 0.4243  & 0.4679 & & 0.5203 & 0.7218 & 0.6047& \\
        \cdashline{2-14}
                             & SVM-P & 0.5661 & 0.6791  & 0.6169 &  & 0.5405 & 0.3742 & 0.4288 &  * & 0.5125 & 0.7373  & 0.6047 &\\
                             & NCWS & 0.5503 & 0.7258  & \textbf{0.6256} & *  & 0.5294 &  0.4890 & \textbf{0.5083}& * & 0.5113 & 0.7421  & \textbf{0.6054} & * \\
        \hline
        \multirow{3}{*}{\textbf{SVM-NoS}} & basic & 0.5597 & 0.6437  & 0.5975 &  & 0.5370 & 0.2603 & 0.3362&  & 0.5268 & 0.6351 & 0.5759 & \\
        \cdashline{2-14}
                             & SVM-P & 0.5418 & 0.6930  & 0.6081 &  & 0.5377 & 0.2421 & 0.3139 & & 0.5268 & 0.6351 & 0.5759 & \\
                             & NCWS & 0.5315 & 0.7330  &\textbf{0.6151}& * & 0.5345 & 0.2799 & \textbf{0.3443} & * & 0.4777 & 0.7468 & \textbf{0.5827} & *\\
        \hline
        \multirow{3}{*}{\textbf{SVM-Structural}} & basic & 0.5819 & 0.6309 & 0.6017 & & 0.5364 & 0.4103  & 0.4576 & & 0.5386 & 0.6805  & 0.6013 &\\
        \cdashline{2-14}
                             & SVM-P & 0.5790 & 0.6378  & 0.6036 &  & 0.5392 & 0.3790 & 0.4343 & & 0.5462 & 0.6621  & 0.5986&\\
                             & NCWS & 0.5501 & 0.7263  & \textbf{0.6252} & * & 0.5267 & 0.4883 & \textbf{0.5011} & * & 0.5169 & 0.7295 & \textbf{0.6051} &  * \\
        \hline
        \multirow{3}{*}{\textbf{SVM-Age}} & basic & 0.5569 & 0.5293 & 0.5428 & & 0.6457 & 0.1986& 0.3037& & 0.6012& 0.4987& 0.5451 &\\
        \cdashline{2-14}
                             & SVM-P & 0.5925 & 0.3990 & 0.4769& & 0.7338& 0.0840& 0.1508& & 0.6598& 0.3336& 0.4431 & \\
                             & NCWS & 0.5159 & 0.7098 & \textbf{0.5975} & * & 0.5910 & 0.2526 & \textbf{0.3539} & * &  0.5939 & 0.5243 & \textbf{0.5569} & * \\
        \hline
        \multirow{3}{*}{\textbf{SVM-ALL}} & basic & 0.6023 & 0.6253 & 0.6136 & & 0.5254 & 0.6023 & 0.5612 & & 0.5975 & 0.6619 & 0.6280 &\\
        \cdashline{2-14}
                             & SVM-P & 0.5651 & 0.6932 & 0.6226 &  & 0.5148 & 0.6237 & 0.5641 & & 0.5472 & 0.6974 & 0.6132 & \\
                             & NCWS & 0.5751 & 0.7063 & \textbf{0.6340} & * & 0.4988 & 0.7152 & \textbf{0.5877} & * & 0.5730 & 0.7086 & \textbf{0.6336} & * \\
        \hline
        \multirow{3}{*}{\textbf{CNN}} & basic & 0.5416 & 0.4674 & 0.5018 & & 0.4969 & 0.4699 & 0.4830 & & 0.4291 & 0.6294 & 0.5103 & \\
        \cdashline{2-14}
                             & SVM-P & - & - & - & & - & - & -& & - & -  & -  &\\
                             & NCWS & 0.5291 & 0.5254 & \textbf{0.5272} & * & 0.4718 & 0.5362 & \textbf{0.5019} & * & 0.4082 & 0.7624 & \textbf{0.5317} & * \\
        \hline      
        \multirow{3}{*}{\textbf{BERT}} & basic & 0.5248	& 0.7338 &  0.6119 & & 0.4783 & 0.6807 & 0.5618 & & 0.5161 & 0.6395 & 0.5712 &  \\
        \cdashline{2-14}
                             & SVM-P & - & - & - & & - & - & -& & - & -  & -  &\\
                             & NCWS & 0.5034 & 0.7927 & \textbf{0.6157} & *	& 0.4543 & 0.7653 & \textbf{0.5701} & * & 0.4923& 0.7127 & \textbf{0.5823} & * \\
        \hline 
\end{tabular}
\end{table*}

\subsection{RQ2: Classification Correction Evaluation}\label{ssec:results:rq2}
For RQ 2, we examine the effectiveness of our NCWS approach along the selected basic classifiers from the previous section's conclusions and in comparison to other existing classification correction approaches (namely SVM-P, C-PU and P-conf) from the literature, on three datasets, namely, the Yelp, Kindle and Electronics datasets. As mentioned in Section~\ref{sec:introduction}, we aim to address the problem of binary classification with incomplete positive instances and abundant unlabelled instances. Following the general weak supervision paradigm, the main objective of our NCWS approach is to help classifiers model the unlabelled instances by identifying further positive instances\footnote{These instances would have otherwise been assumed to be negative.} from the many unlabelled instances, thereby improving their performances.

\looseness -1 \wang{First, we compare the differences between the classification results of the basic classifiers and the results after applying the corresponding classification correction approaches \xwang{to the basic classifiers}. \craig{Figures
~\ref{fig:baselines}} \craig{plots} the frequency distributions of the predicted scores of review instances \craig{(in \yaya{the} range -1 to 1)}, to illustrate the alteration of \yaya{the} classification results \yaya{when NCWS and the classification correction baseline approaches are deployed}. 
\craig{For reasons of brevity,} we \yaya{show} the classification results of the SVM-ALL classifier, \yaya{the} \yaya{best performing} classifier, as a representative example \yaya{ of the effects of applying NCWS and the various classification correction methods on the Kindle dataset}. \yaya{However, the observed trends remain consistent \yaya{across all basic classifiers} for the other 2 Yelp and Electronics datasets}.
\yaya{In particular}, Figure~\ref{fig:basic} shows the \yaya{review helpfulness score predictions} given by the basic \xwang{SVM-ALL} \yaya{classifier}, \yaya{while Figure~\ref{fig:NCWS} shows the  \yaya{obtained predictions} after correcting the classifier using} our NCWS approach. It is clear from the figures that when NCWS is applied, a further number of reviews are classified as positive instances by the corrected SVM-ALL classifier.}
\pageenlarge{2} \yaya{These results} are in line with the objective of our NCWS approach to identify more positive examples from \yaya{the} unlabelled instances. \yaya{\xwang{Figures 2(c)-(e) show that after applying the classification correction baseline approaches (i.e.\ SVM-P, C-PU and P-conf) \yaya{to} the SVM-ALL classifier, the helpfulness score \yaya{prediction} distributions of \yaya{the} reviews \yaya{change} in different ways.}}
\yaya{While SVM-P \yaya{adjusts} the classification results of the basic \xwang{SVM-ALL} classifier by identifying further positive review instances, C-PU and P-conf completely change the \xwang{frequency} distribution of the original \xwang{SVM-ALL} \yaya{classifier's score predictions}.} In particular, they classify most review instances as positive with C-PU \craig{squeezing most} of the classification results \craig{into} an extremely narrow range (i.e.\ between 0.9994 and 0.9998). As a consequence, only NCWS and SVM-P \yaya{are useful at} \yaya{correcting and enhancing the performance of the SVM-ALL} classifier. \yaya{Hence, in the following, we focus our experiments on the best \craig{performing} basic classifiers -- \craig{as identified in the conclusions of RQ1} -- and the corresponding NCWS and SVM-P corrected results, on the 3 datasets, to further examine their overall effectiveness}. 

\looseness -1 \wang{\yaya{Table~\ref{tab:correction_effect} examines} the effectiveness of SVM-P and NCWS after applying them to the SVM and NN-based classifiers on the 3 datasets. \craig{For} the SVM-P approach, the F1 scores show that SVM-P can be helpful to 
all the basic classifiers \xwang{-- except SVM-Age --}
and \yaya{enhances} their classification performance with higher F1 scores. However, such effectiveness is not generalisable to all SVM-based approaches on the 3 datasets. For example, SVM-P negatively \yaya{impacts} the SVM-LEN classifier on the Kindle dataset with a lower F1 score (0.4679 \craig{$\rightarrow$} 0.4288). Meanwhile, our NCWS approach outperforms SVM-P by \craig{exhibiting} higher F1 scores. In particular, it can significantly and consistently enhance the basic SVM-based classifiers \yaya{to yield} higher F1 scores \xwang{according to the McNemar’s test.}
On the other hand, apart from these SVM-based classifiers, we further examine the effectiveness of NCWS\footnote{\wang{The SVM-P \yaya{method} is limited to SVM-based classifiers as introduced in Section~\ref{ssec:classifier_exp}.}} on the NN-based \yaya{classifiers}, which have \yaya{been} shown to be effective in many neural language processing tasks. \yaya{Table~\ref{tab:correction_effect} shows that significant and consistent improvements are observed} after applying NCWS to \yaya{CNN and BERT-based NN classifiers on the 3 datasets}. This \yaya{demonstrates} the effectiveness of NCWS \xwang{when \craig{applied}} to the cross-entropy loss function. Moreover, \yaya{when applied to the \yaya{best performing} approaches} \xwang{(i.e.\ SVM-LEN, SVM-Structural, SVM-ALL and BERT)},
NCWS can further \yaya{increase the F1 scores of} \yaya{these} classifiers. \pageenlarge{2} For example, NCWS \yaya{improves} SVM-ALL to obtain the \yaya{overall} best classification performance \yaya{on the 3 datasets}.}

\begin{table}[tb]
    \centering\scriptsize
    \caption{\ya{NCWS's impact on classification: number of unlabelled reviews that changed from being predicted negative to predicted positive by the application of NCWS; total number of reviews predicted negative by the basic classifiers; the percentage of reviews that changed is also shown.}
    }\vspace{-2mm}
    \resizebox{85mm}{!}{
    \setlength\tabcolsep{1.5pt}
    \begin{tabular}{|l|cc|cc|cc|}
    \hline
         & \multicolumn{2}{c}{\textbf{Yelp}} & \multicolumn{2}{c}{\textbf{Kindle}} & \multicolumn{2}{c|}{\textbf{Electronics}}\\
    \hline
        \textbf{SVM-LEN} & 7559 / 90260 & (8.3\%) & 10030 / 92138 &(10.8\%) & 4587 / 126187 &(3.6\%)\\
        \textbf{SVM-NoS} & 15714 / 97301& (16.1\%) & 2563 / 104016& (2.4\%) & 31255 / 138253 &(22.6\%)\\
        \textbf{SVM-Structural} & 20534 / 95746& (21.4\%)& 6888 / 93687& (7.3\%)& 12562 / 136656& (9.1\%)\\
        \textbf{SVM-Age} & 52797 / 160458 &(32.9\%)& 9781 / 171451 &(5.7\%)& 6778 / 232405 &(2.9\%)\\
        \textbf{SVM-ALL} & 23603 / 149574& (15.78\%)& 20235 / 101828& (19.8\%)& 21413 / 201927& (10.6\%)\\
        \textbf{CNN} & 25204 / 171307 &(14.7\%) & 35012 / 119640 &(29.3\%) & 50397 / 152453& (33.1\%)\\
        \textbf{BERT} & 12503 / 78451& (15.9\%) & 9217 / 68815 &(13.3\%)& 10512 / 165227& (6.3\%)\\
    \hline
    \end{tabular}}
    \label{tab:correction}\vspace{-4mm}
\end{table}

\looseness -1 \yaya{Table~\ref{tab:correction} shows how many unlabelled reviews were classified by the basic classifiers as negative \craig{but classified} as positive by our NCWS approach}. For example, in the Yelp dataset, the \ya{SVM-LEN basic classifier predicts} 90,260 unlabelled examples as unhelpful while NCWS \ya{predicts} 7,559 instances of these reviews as being helpful (i.e.\ \ya{an $\sim$8.3\% increase}). In general, about 5-30\% of the unlabelled reviews are re-labelled as positive and identified as helpful by NCWS. These results align with our objective to identify more positive instances from the unlabelled corpus. Therefore, by analysing the results from the three datasets, \wang{we \yaya{can now} answer RQ 2: NCWS can successfully improve the performances of basic classifiers while outperforming other classification correction approaches on the F1 evaluation metric.} These results validate our assumption that the age of reviews is a reliable signal to infer which of the unlabelled reviews are actually unhelpful and that the longer an unlabelled review has been posted, the more likely this review is unhelpful. 
In the next section, we \yaya{show} how \yaya{the performance of a state-of-the-art} review-based venue recommendation recommendation system can benefit from the further identified positive instances.

\section{Venue Recommendation Application}\label{sec:rq3}

\pageenlarge{2}\looseness -1 In this section, we use venue recommendation as a use case to demonstrate the benefit of NCWS, thereby addressing RQ 3 stated in Section~\ref{ssec:research_questions}, which focuses on examining the usefulness of the identified helpful reviews in enhancing \yaya{the} review-based recommendation performance. Venue recommendation is an important application of recommendation techniques, where the aim is to suggest relevant venues that a user would like to visit. Often, venue recommendation approaches make use of data from a location-based social network such as Foursquare or Yelp. This data can be implicit feedback, in the form of checkins, or more explicit, such as ratings or reviews~\cite{wang2019comparison}.

We apply NCWS in the context of a state-of-the-art venue recommendation model, namely DeepCoNN~\cite{zheng2017joint}, which uses the reviews of users on venues for recommendation as input to a convolutional neural network. DeepCoNN constructs two parallel convolutional neural networks to model user and venue reviews, respectively, to predict the rating that a user would give to a venue. Indeed, DeepCoNN has been frequently used in many review-based recommendation works as a baseline~\cite{chen2018neural,rafailidis2019adversarial}.  

To integrate NCWS into DeepCoNN, we simply replace the user and venue review corpora with only those reviews that are predicted to be helpful. In doing so, we postulate that removing noisy or unlabelled reviews and replacing them with further positive reviews as identified by NCWS results into a more effective learned DeepCoNN model.  We validate this through experiments on the Yelp dataset, which is a widely used venue recommendation dataset~\cite{wang2019comparison}.

\subsection{Experimental Setup}
 \looseness -1 We compare different replacement strategies to assess the effectiveness of \xiwang{the corresponding review selection approaches as input for DeepCoNN.} Such review selection approaches include: (1) \textbf{\xiwang{+Random:} } randomly samples the same numbers of reviews as the predicted helpful reviews with the best corrected classifier \wang{(i.e.\ SVM-ALL)} \craig{as input for DeepCoNN}; 
(2) \textbf{\xiwang{+Basic:} } selects helpful reviews 
with the basic \wang{SVM-ALL} classifier, which had the best performance in review helpfulness classification in Section~\ref{ssec:results:rq1}; (3) \textbf{\xiwang{+NCWS:} } uses the predicted helpful reviews with a NCWS-corrected \wang{SVM-ALL} classifier. 
\xwang{Similarly, \yaya{as additional comparative approaches,} we use the corresponding} 
\xwang{(4) \textbf{\xiwang{+SVM-P}}, (5) \textbf{\xiwang{+C-PU}} and (6) \textbf{\xiwang{+P-conf}} but with the SVM-P, C-PU and \yaya{P-conf's} predicted helpful reviews instead.} 
We apply these different review-selection strategies within the DeepCoNN venue recommendation model on the Yelp dataset, along with two baselines: DeepCoNN and \yaya{a} popular rating prediction approach, namely Non-negative Matrix Factorisation (NMF)~\cite{sra2006generalized}, \yaya{which} considers ratings, \yaya{but} not the text of the reviews.

\pageenlarge{2}\looseness -1 Our experiments are conducted using a 5-fold cross validation, following as closely as possible the experimental setup of~\cite{zheng2017joint} (with the same trained word embedding model, but with a different dataset and review selection strategies). We evaluate the rating prediction accuracy using Mean Average Error (MAE) and Root Mean Square Error (RMSE). For both metrics, smaller values are better. We use the paired t-test to determine significant differences of MAE\footnote{RMSE, which is a non-linear aggregation of squared absolute errors, is not suitable for significance testing\ya{.}}.

\subsection{Results}

\begin{table}[tb]
    \centering\scriptsize
    \caption{\looseness -1 Venue recommendation results: Significant \craig{MAE improvements} (t-test, $p$$<$$0.05$) \xiwang{w.r.t.\ }DeepCoNN, \xiwang{DeepCo\-NN+Basic} \& \xiwang{DeepCoNN+SVM-P} are \craig{denoted} by $\circ$, $\bullet$ \& $*$, resp..}\vspace{-2mm}
    \label{tab:venue_recommendation}
    \resizebox{55mm}{!}{\begin{tabular}{|l|c|c|}
    \hline
                 & MAE & RMSE \\
    \hline
     NMF & 1.1526 & 1.4345\\
    \hdashline
    DeepCoNN & 0.8969 & 1.1798\\
    \hdashline
    \xiwang{+Random} & 0.9201 $\circ *$ & 1.2278\\
    \xiwang{+Basic} & 0.8629 $\circ$ & 1.1012  \\
    \xiwang{+C-PU} & 0.8969 $\bullet *$& 1.1798  \\
    \xiwang{+P-conf} & 0.8969 $\bullet *$& 1.1798  \\
    \xiwang{+SVM-P} & 0.8597 $\circ$ & 1.0998 \\
    \xiwang{+NCWS} & \textbf{0.8503 $\circ \bullet *$} & \textbf{1.0954} \\
    \hline
    \end{tabular}} \vspace{-3mm}
\end{table}

\looseness -1 Table~\ref{tab:venue_recommendation} presents the MAE and RMSE scores of the DeepCoNN variants. In particular, the first group of rows are baselines, while the second group \yaya{corresponds} to approaches that make use of review helpfulness classification when filtering the set of reviews to use. On analysing the table, comparing the rating \xwang{prediction error \yaya{of DeepCoNN and NMF} \yaya{using} the MAE and RMSE \yaya{metrics}}, we observe 
that \yaya{DeepCoNN, which uses \xiwang{all} reviews} for venue recommendation enables better representations of user preferences and venue properties with \xwang{lower rating prediction \yaya{errors} \yaya{than NMF}}. Indeed, recall that NMF is only trained on ratings, while DeepCoNN has access to the text of the reviews. In relation to the helpful reviews, we observe that by filtering \craig{the} reviews to include only those that are predicted to be helpful (c.f.\  DeepCoNN\xiwang{+Basic}\xiwang{, +SVM-P and +NCWS}), the rating prediction is improved (i.e.\ significantly reduced MAE \yaya{and} RMSE \yaya{scores}) compared to DeepCoNN. This implies that, among all the reviews considered \craig{by the} DeepCoNN \craig{baseline}, some are noisy, and removing these to focus upon the likely helpful reviews \craig{aids learning} an effective venue recommendation model. \craig{On the other hand, the +C-PU and +P-conf integrations have the same performances as DeepCoNN -- indeed, this is expected from the results of Section 6.2, where P-Conf and C-PU were not effective in identifying helpful reviews.}


Moreover, comparing the performances between \craig{the} \xiwang{+Basic, +SVM-P and +NCWS} \craig{integrations}, we find that our proposed NCWS approach results in better rating predictions, exhibiting a significant 1.8\%  improvement in MAE and 0.27\% improvement in RMSE, respectively, in comparison to the \xiwang{+Basic} approach that uses the basic \xwang{SVM-ALL} \yaya{classifier}. \xiwang{In particular, \xiwang{+NCWS} also shows \yaya{a} significantly better performance than \yaya{\xiwang{+SVM-P}}, which indicates the benefit of using NCWS \yaya{over} SVM-P \yaya{in identifying} helpful reviews \craig{as input for} the DeepCoNN model.} 
Moreover, \xiwang{+NCWS} is significantly more accurate than \xiwang{+Random}, a variant that uses the same number of randomly sampled reviews. Hence, and in answer to RQ3, we find that focusing on the likely helpful reviews, particularly those additional reviews found using our proposed NCWS classifier (\craig{see} Table~\ref{tab:correction}), allows the 
\xwang{performance} 
of the state-of-the-art DeepCoNN rating prediction approach to be significantly enhanced. This also validates the effectiveness of NCWS in identifying helpful reviews.

\section{Conclusions}\label{sec:conclusion}
\pageenlarge{2}
\looseness -1 In this paper, we proposed a \yaya{novel} weak supervised binary classification correction approach by considering the negative confidence of \yaya{the} unlabelled examples under the positive and unlabelled learning scenario. Using three datasets, we showed the effectiveness of \yaya{our NCWS approach} in comparison to several state-of-the-art classification correction approaches from the literature. We also illustrated how NCWS allows to increase the number of positive instances by \wang{5--30}\% when integrated into various binary classifiers. NCWS is a general classification correction approach, which can be applied to various \yaya{other} classification tasks for which a negativity score can be defined. Using review helpfulness classification as a use case, we extensively demonstrated the effectiveness of NCWS in leveraging the predicted helpful reviews to significantly enhance the performance of DeepCoNN, a recent and strong review-based recommendation model. \ya{As} future work, we plan to apply NCWS to other binary classification use cases where there are limited labelled examples.

\bibliographystyle{ACM-Reference-Format}
\bibliography{sample-base}
\end{document}